# Helix formation in linear achiral dendronized polymers. A computer simulation study.

D.K. Christopoulos[(1)], A.F. Terzis[(2),*], A.G. Vanakaras[(1)] and D.J. Photinos[(1)]

[(1)] *University of Patras, Department of Materials Science, Patras 26504, Greece*

[(2)] *University of Patras, Department of Physics, Patras 26504, Greece*

**Abstract**

We present a molecular simulation study of the structure of linear dendronized polymers. We use excluded volume interactions in the context of a generic coarse-grained molecular model whose geometrical parameters are tuned to represent a poly(para-phenylene) backbone with benzyl ether, Fréchet type dendrons. We apply Monte Carlo sampling in order to investigate the formation of packing-induced chiral structures along the polymer backbone of these chemically non-chiral systems. We find that helical structures can be formed, usually with defects consisting of domains with reversed helical handedness. Clear signs of helical arrangements of the dendrons begin to appear for dendritic generation *g=4*, while for *g=5* these arrangements dominate and perfect helices can even be observed as equilibrium structures obtained from certain types of starting configurations.

* Author to whom correspondence should be addressed. Fax: +30 2610 997618; Electronic mail: terzis@physics.upatras.gr

## I. INTRODUCTION

Linear dendronized polymers (LDPs) are a special class of nanoscopic molecules that have recently attracted increasing scientific attention due to their unique structure, properties and potential applications.[1] Commonly, LDPs consist of a linear polymeric core (the "backbone") onto which dendritic units (the dendra) are grafted at regular intervals.[1] A key feature of LDPs is the coexistence, in a single molecular structure, of three distinct topological regions: the internal backbone, the dendron region around the backbone, and an external surface, usually amenable to functionalisation. The high controllability of the LDP structures stems from the possibility of controlling the generation and the grafting density of the dendrons as well as modifying chemically the external surface.[2] LDPs are currently under intense investigation with respect to various applications, including the synthesis of hierarchically structured materials, catalysis, optoelectronic applications, as well as applications in the biosciences, such as ion channel mimics and DNA compactization.[3] Understanding the principles that govern the nanomechanics of high molecular weight single molecules, such as dendrimers and dendronized polymers, is a scientific challenge in its own right and is moreover instrumental to the successful fabrication of single molecule devices.

While for simple dendrimers, the generation of the dendrons, the branch size and the coordination number (number of connected branches at each branching point) are the essential structural parameters, for LDPs, the grafting density of the dendra along the backbone contour is also an important parameter.[1-5] The optical,[6] dymamic[7] and conformational properties[8] of LDPs depend strongly on the architecture, on the generation and on the grafting density of the side dendrons.[9]

It is known experimentally that linear dendronized polymers may form helical structures when the backbone and/or the attached dendrons are chiral.[10] The formation of these helical structures depends on the generation of the grafted dendrons and examples are known of both, high generations favouring[11] or disfavouring[12] such structures. In the later situation the polymer backbone is capable of forming $a$-helices in the absence of grafted dendrons, a structure that persists for the low-generation dendronized backbones and undergoes a conformational transition to non helical structures upon increasing the dendron generation. Furthermore, formation of double-stranded fibers with well defined diameters has been observed experimentally[13] and is attributed to the hydrophobic effect and to crowding in the dendron shell of the third and fourth generation LDPs.

The present work addresses the formation of chiral structures in intrinsically non-chiral LDPs. Confinement and excluded volume interactions (packing) may lead to the spontaneous appearance of chiral asymmetries. This is directly verified by theoretical investigations,[14] and subsequent experimental confirmations thereof,[15] showing that cylindrical confinement of sterically interacting spherical particles may lead to the formation of macroscopically chiral structures when the particle diameter is comparable to the diameter of the confining cylinder. High generation LDPs are endowed with all the essential features presented by the system of cylinder-confined spheres: (a) The geometry of the system is cylindrical, at least on a length scale of a few tens of nanometers for a relatively rigid backbone, (b) the laterally attached dendrons are restricted to move about the backbone axis within a well defined, generation-dependent, radius and (c) short range, excluded volume interactions between dendrons are expected to dominate the local structure of the LDP, although specific interactions between submolecular units (e.g. multipolar electrostatic interactions, hydrogen bonding, amphiphilic interactions between the LDP and its environment) could influence its conformational thermodynamics .

The theoretical study of single LDPs by molecular simulation methods,[4,5,16] pose certain difficulties which stem from the enormous number of accessible conformations and of the branching topology of these molecules. As various length and time scales are involved in the statistical mechanics of LDPs and the molecular weight grows exponentially with the generation of the grafted dendra, the use of coarse grained models becomes inevitable. Moreover, even when coarse-grained models are employed, reliable macromolecular structural properties of high generation and dense grafting LDPs require very long simulations due to slow equilibration.[4] According to recent experimental[11,17] and computational[17] results, this slow equilibration is an inherent property associated with the "glassy" dynamics of the high generation LDPs. As a result, molecular simulations of these macromolecular systems require special treatment in order to obtain the correct equilibrium (or near equilibrium) properties.

In this work we present simulation results on the spontaneous helix formation in a generic model LDP that consist of a non-chiral backbone decorated with achiral dendrons. The LDP is taken to consists of a relatively stiff backbone with structural properties similar to those of the poly-para-phenylene molecule. This backobone is decorated with benzyl ether, Fréchet type dendrons on every second phenylene ring.[18-20] We study the structural properties of the system as a function of the generation of the grafted dendrons. Neglecting details on the atomistic scale as well as specific interactions either between dendrons or between the LDP and its environment, we investigate how the peripheral crowding may lead to a spontaneous helical ordering in high generation LDPs that renders them, at least locally, chiral.

The paper is organized as follows. In Sec. II we present the coarse grained representation used for the modeling of the LDPs, together with details on the simulation method. In Sec. III we present simulation results revealing the formation of helical structures along the LDPs and we study the influence of the dendritic generation on these structures. In Sec. IV we consider simulation results regarding properties of the LDPs that do not reflect directly the formation of helical structures and we show that these results are consistent with the presence of helical structures and more specifically with the onset of such structures at high dendritic generation. The main conclusions of the present study are stated and discussed in Sec. V.

## II. MODEL DESCRIPTION AND SIMULATION DETAILS

Consider a LDP of the generic structure shown in Fig. 1(a). It consists of a poly(para-phenylene) (PPP) backbone with benzyl ether, Fréchet type dendra.[1,18-20] In a coarse grained representation, Fig. 1(b), the LDP is modeled as a covalent assembly of spherical united atoms, or sites, of diameter $D = 6.34$ Å, corresponding to the effective Van der Waals diameter of a phenyl ring. The number of such sites in a single repeat-unit of generation $g$ is $N_{dendron}(g) = 1 + 2^{g+1}$. The conformational states of the LDP molecule are labeled by the index $n$. To keep the molecular model as simple as possible, retaining at the same time the essential characteristics of the real molecule, we have assumed the following form for the conformational (*intra*-molecular) potential energy of the LPD in conformation $n$,

$$V(n) = \sum_{\substack{bonded \\ pairs\{s,s'\}}} u^{(BL)}(l_{ss'}) + \sum_{\substack{backbone \\ bond\, angles\{\vartheta_i\}}} u^{(BA)}(\vartheta_i) + \sum_{\substack{non-bonded \\ pairs\{s,s'\}}} u^{(NB)}(r_{ss'}) \quad (1)$$

Here $u^{(BL)}(l_{ss'})$ is the potential describing bonded pairs of sites and depends on the length $l_{ss'}$ of the bond connecting the sites $s$ and $s'$. The summation extends over all possible pairs of bonded sites, i.e. successive sites on the backbone, adjacent sites on the same dendron and backobone-dendron sites jointed by grafting bonds. In our calculations it is assumed that the length of the bond connecting successive backbone sites varies freely in the range $(0.67 \pm 0.09)D$. That is, $u^{(BL)}(l_{ss'})$ is taken to vanish if $l_{ss'}$ lies within this range and to become infinitely large otherwise. Numerically, this range is suggested by simple molecular mechanics calculations[21] for the centre-to-centre distance between two successive phenyl rings of a single

PPP molecule for a wide range of temperatures. Similarly, bonds connecting sites on the same dendron are taken to vary freely in the length range $(1.02 \pm 0.02)D$. The same range is also used for bonds connecting dendron sites to backbone grafting sites.

The second term in Eq. (1), $u^{(BA)}(\vartheta_i)$, is the bond angle deformation potential and refers only to the PPP backbone, with $\vartheta_i$ denoting the angle between two successive bonds of the backbone (see Fig. 2). $u^{(BA)}(\vartheta_i)$ is assumed to vanish if $\theta < \pi/6$ and to become infinitely large otherwise. This confers reasonable bending flexibility to the backbone, in accord with Molecular Mechanics calculations,[21] while maintaining the local rigidity of the biphenyl.

Finally, the $u^{(NB)}(r_{ss'})$ terms in Eq. (1) describe the interaction between non-bonded sites $s, s'$ of the LDP and are assumed to depend only on the site-site distance $r_{ss'}$. All the quantitative results presented in this work are obtained assuming hard body interactions between the sites, with the contact distance taken equal to the diameter $D$ of the sites.

For thermal equilibrium at temperature $T$, the probability $p(n)$ for the supermolecule to be found in the conformational state $n$ is given by

$$p(n) = \exp[-V(n)/k_B T] / \sum_{\{n'\}} \exp[-V(n')/k_B T], \qquad (2)$$

where $k_B$ is the Boltzmann constant and the summation in the denominator runs over all the conformational states of the supermolecule. The standard Monte Carlo (MC) method[22] is applied in order to sample the conformational-phase-space with the probability distribution given in Eq. (2). A MC move consists of a random displacement of a randomly selected single site. The length of the displacement is adjusted during the simulation so as to yield a value of about 0.2 for the ratio of the accepted over the attempted displacements. The number of attempted moves in a MC cycle is, on average, ten times the total number of sites of each simulated LDP. A typical run consists of $10^6$ Monte Carlo cycles for equilibration followed by $10^5$-$10^6$ cycles during which structural data are collected. Accordingly, the total simulation time of our calculations grows exponentially with dendron generation.

The output of the present simulations is a set of probability distribution functions and average values for various physical quantities (observables) that describe the equilibrium structure of the LDPs. These observables are described in Ref. 4 together with details concerning the construction of overlap-free initial configurations of the supermolecule. All the results presented in this paper are for model LDPs comprising $N$=20 monomers and for dendron generations from $g = 0$ up to $g = 5$. Additionally, simulations were performed for LDPs with $N$=10, 15 and $N$=30 monomers in order to explore the system size effects on the calculated quantities. These simulations indicated that the calculated conformational averages presented in this paper show marginal variation for $N$ above 15.

## III. RESULTS ON HELICAL STRUCTURES

For each pair of consecutive dendra along the backbone of the LDP, a torsion angle $\phi_i$ is defined in terms of the three unit vectors $\mathbf{u}_i, \mathbf{e}_i, \mathbf{u}_{i+1}$ shown in Fig. 2. The unit vector $\mathbf{e}_i$ is on the line connecting the $i$ and $i+1$ grafting sites of the backbone and the two unit vectors $\mathbf{u}_i$, $\mathbf{u}_{i+1}$ are along the directions connecting these sites with the centers of mass of the grafted dendra $i$ and $i+1$ respectively. The value of $\phi_i$ is given by the magnitude of the dihedral angle formed by the planes of the vectors $(\mathbf{u}_i, \mathbf{e}_i)$ and $(\mathbf{e}_i, \mathbf{u}_{i+1})$, with $\phi_i$=0 taken to correspond to the planar "parallel" disposition of the vectors $\mathbf{u}_i, \mathbf{u}_{i+1}$ (i.e. $\mathbf{u}_i \cdot \mathbf{u}_{i+1} > 0$). Accordingly, the physically coincident values $\phi_i = \pm 180°$ corresponds to the "antiparallel" disposition ($\mathbf{u}_i \cdot \mathbf{u}_{i+1} < 0$) depicted in Fig. 2.

We have calculated the local probability distribution functions $p_i^{(t)}(\phi_i; g)$ of the torsion angle $\phi_i$. This set of torsion probability distributions serves as a direct indicator of the occurrence of systematic twist along the LDP. The arithmetic average of this set along the entire backbone yields the mean probability distribution function of the torsion angles for the LDP,

$$p^{(t)}(\phi; g) = \frac{1}{N-1} \sum_{i=1}^{N-1} p_i^{(t)}(\phi; g) \quad . \tag{3}$$

The results of our calculations are shown in Fig. 3. It is apparent from the plots 3(a)-(b) that for the two lowest generation ($g$=0,1) LDPs the torsional angle samples all possible values with substantial probability. This probability is lowest in the region around $\phi = 0°$, which corresponds to the parallel disposition of the **u** vectors of the successive dendra (which brings the outer shells of these dendra at the closest mutual distance). On going to $g$=2 (Fig. 3(c)), the low probability region becomes deeper and broader, while two local maxima of the probability appear around $\pm 120°$. This trend is further enhanced for generation $g$=3, where, as seen in Fig. 3(d), the local maxima evolve into peaks while the probability for torsion angles around $\phi = 0°$ is low but not negligible. The peaks are positioned nearly symmetrically at $\pm 120°$ but their heights are not symmetric, indicating that, in this particular sample, the populations of positive and negative torsions are not equal. However, the accessibility of all possible torsion angles allows for reasonably rapid changes in these populations. This makes it possible for a distribution like the one shown in Fig. 3(d) to evolve, on prolonging the simulation, into a distribution with inverted relative heights of the $\pm 120°$ peaks. In very long simulations, the system can be observed to fluctuate back and forth between states of opposite dominant torsion.

The $g$=4 LDPs clearly exhibit a bimodal distribution with asymmetric peaks centered at $\phi \approx \pm 120°$ and with practically vanishing probabilities in the region around $\phi = 0°$. In the particular instance shown in Fig. 3(e), the structure is dominated by the $\phi \approx +120°$ torsions, corresponding to a helix whose pitch spans three repeat units, but there is a considerable number of $\phi \approx -120°$ "defects" in this structure. Again, in the course of a longer simulation, such a distribution can evolve into one with the relative heights of the peaks inverted, corresponding to a helical structure dominated by $\phi \approx -120°$ torsions interrupted by $\phi \approx +120°$ "defects". In this case, however, the evolution is much slower than for the $g$=3 LDPs because (a) the range around $\phi = \pm 180°$, which is the main and shortest path for the evolution of a $\phi \approx +120°$ torsion into one with $\phi \approx -120°$ and *vise versa*, has much lower probability than for the $g$=3 LDPs and (b) the range of torsion angles around $\phi = 0°$ is practically inaccessible. A closer analysis of the LDP structures associated with the distribution of Fig. 3(e) indicates that the shape of this distribution is a result of a statistical mixture of two kinds of helical structures, each of which presents a number of defects that interrupt the dominant succession of torsions with torsions of the opposite sense.

The situation is quite different for the $g$=5 LDP: The peaks of the distribution are centered around $\phi \approx \pm 90°$, they are much sharper than those of $g$=4 and the distribution practically vanishes in the rest of the torsion angle range. If the starting conditions are such as to favour only the positive or the negative torsions then only one peak appears in the equilibrated system (the $+90°$ or the $-90°$ respectively). In the instance of Fig. 3(f), there is a single peak centered $-90°$. This structure corresponds to a defect-free helix with a pitch comprising four repeat units of the LDP. Furthermore, this structure is stable over rather lengthy simulation runs, in accord with the inaccessibility of torsion angles outside the region spanning the breadth of the single peak. On the other hand, if the starting conditions favour the creation of some defects in the helical structure of the LDP, then these defects appear to persist over very long runs suggesting that the system is in these cases trapped in local minima of the free energy. The very long persistence of these defects is attributed to the use of hard body interactions. This is in accord with the results of test simulations in which the interactions are gradually changed to soft repulsions, yielding progressively less persistent defects.

A representative snapshot obtained for the $g=5$ LDP in the course of the simulation leading to the distribution of Fig. 3(f) is shown in Fig. 4. The formation of the helical structure is clearly depicted and a microscopic interpretation of its stability is suggested: The helical arrangement is the most favorable way to pack these bulky dendra under the constraints imposed by their grafting to the backbone and once this arrangement is reached the packing barrier for defects or for the transition to a defect-free structure of the opposite helical sense is rather high. All the quantitative results to be reported subsequently for the g=5 LDPs correspond to the 'defect-free' helical state associated with the plot of Fig. 3(f).

The question may now be posed of whether this twisting of the dendra is accompanied by some curling of the backbone. To answer this question we have studied the angular distribution of successive $\mathbf{e}_i$ the unit vectors (see Fig. 2) of the backbone. This involves two sets of angles, namely (a) the "bending" angles $\theta_i$ between two successive unit vectors, defined by $\cos\theta_i = \mathbf{e}_i \cdot \mathbf{e}_{i+1}$, and (b) the dihedral angles $\psi_i$ defined by three unit vectors ($\mathbf{e}_{i-1}, \mathbf{e}_i, \mathbf{e}_{i+1}$) that connect successive grafting sites of the backbone. Obviously, a dihedral angle $\psi_i$ is physically relevant only if none of the bending angles $\theta_{i-1}$ and $\theta_i$ is too close to a vanishing value.

We have calculated the average probability distribution $p^{(b)}(\cos\theta; g)$ of the bending angles over the entire backbone, defined according to

$$p^{(b)}(\cos\theta; g) = \frac{1}{N-2}\sum_{i=1}^{N-2}\left\langle \delta\left(\cos\theta - \mathbf{e}_i \cdot \mathbf{e}_{i+1}\right)\right\rangle. \qquad (4)$$

The results on $p^{(b)}(\cos\theta; g)$ for $g=0$ to 5 are plotted in Fig. 5(a) and they indicate that, as the generation $g$ increases, the peaks of the distributions progressively move away from the $\cos\theta = 1$ value, i.e. the probability for collinear backbone segments decreases. Nevertheless, as shown in Fig. 5(b), the average values $<\cos\theta>$ calculated from these distributions increase with increasing generation $g$, which indicates that the large angle tails of the distributions diminish as well, with increasing $g$.

Regarding the dihedral angles $\psi_i$, the important point of the results in Fig. (5) is that, at least for the g=5 LDPs, there is only a small probability of collinearity of successive unit vectors $\mathbf{e}_{i-1}, \mathbf{e}_i, \mathbf{e}_{i+1}$ and therefore each local distribution function $p_i^{(c)}(\psi_i, g)$ of the dihedral angles is a physically relevant indicator of the possible curling of the backbone. In analogy with the torsion probability distributions of Eq. (3), we have calculated the average of the set of $p_i^{(c)}(\psi_i; g)$ distributions along the backbone to obtain the mean probability distribution $p_i^{(c)}(\psi_i; g)$ function for "backbone curling" of the LDP, i.e.

$$p^{(c)}(\psi; g) = \frac{1}{N-2}\sum_{i=2}^{N-1} p_i^{(c)}(\psi; g) \quad . \qquad (5)$$

The results of our calculations for these distributions are presented in Fig. 6 for $g=2$ and $g=5$. The $\psi$ distribution for the second generation LDP is plotted in Fig. 6(a) and is qualitatively representative of the respective distributions of the $g=0,1$ and 3 LDPs: an essentially flat distribution with a very broad and shallow minimum around $\psi = 0°$. In contrast, the distribution for the $g=5$ LDP, plotted in Fig. 6(b), is essentially a single peak distribution around $\psi = 90°$. This shows that the helix formation in these LDPs involves both, the twisting of the dendra around the backbone and some curling of the backbone, with equal pitch to the twisting. It may be noted that the peak in the $\psi$ distribution (backbone curl) of Fig. 6(b) shows considerably more spread than the peak in the $\phi$ distribution (dendron twist) of Fig. 3(f). This is related to the fact that in the latter case the unit vectors $\mathbf{u}_i$, $\mathbf{u}_{i+1}$ are essentially at right angles to the $\mathbf{e}_i$ vectors (see Fig. 2), thus providing sharply defined dihedral angles $\phi_i$, whereas the relatively small angles formed by the $\mathbf{e}_{i-1}, \mathbf{e}_i, \mathbf{e}_{i+1}$ vectors lead to some smearing of the values of the dihedral angles $\psi_i$.

An alternative way to monitor the formation of helical structures is offered by the study of the mass distribution of the dendra along the LDP. This is done with the help of a set of topologically specific radial correlation functions between sites belonging to different dendrons. These correlation functions are defined according to:

$$\eta_m^{g_t,g_t'}(r) \equiv \frac{1}{2N_{g_t} N_{g_t'}} \left\langle \sum_{s,s',i} \left[ \delta\left(r - \left|\mathbf{r}_{s(i;g_t)} - \mathbf{r}_{s'(i+m;g_t')}\right|\right) + \delta\left(r - \left|\mathbf{r}_{s(i;g_t')} - \mathbf{r}_{s'(i+m;g_t)}\right|\right) \right] \right\rangle, \quad (6)$$

and give the probability of finding a pair of segments at distance $r$, under the constraints that these segments belong to the topological shells of generation $g_t$ and $g_t'$ of two dendrons whose grafting sites are $m$ monomeric units apart. In Eq. (6), a site index of the type $s(i;g_t)$ refers to a site belonging to the topological shell of generation $g_t$ of a dendron that is grafted onto the $i^{th}$ grafting site of the backbone. $N_{g_t}$ denotes the total number of sites in the LDP that belong to a topological shell of generation $g_t$. The topologically specific pair correlation functions are useful indicators of the propagation of possible structures associated with specific relative dispositions of pairs of dendra as a function of their separation along the LDP. We have calculated such topologically specific correlation functions for the outer shells of the LDPs of generations $g=1$ to 5.

Of particular interest, in connection with the helical structures, are those of generations $g=4$ and 5. The respective results are shown in Fig. 7, together with the results of $g=3$ for comparison, where we plot the radial correlation functions $\eta_m^{g_t,g_t'}(r)$ for pairs of sites belonging to the outer topological shells ($g_t = g_t' = g$) of dendra grafted at $m^{th}$ neighbour positions along the backbone (in the notation of Eq. (6) these correlation functions would be denoted as $\eta_m^{g,g}(r)$). The $m$ dependence of these functions is indicative of persistent structures along the backbone. In particular, while the grafting distance of any reference dendron from its $m^{th}$ neighbour dendron is proportional to the index $m$, the distance of the sites at the outer topological shell of that dendron from the respective sites of its $m^{th}$ neighbour dendron could deviate from this proportionality, or even not follow an ascending dependence on $m$, if the dendra present a twisted arrangement along the backbone. This appears to be the case for the $m$ dependence of the $\eta_m^{g,g}(r)$ at high generation $g$.

Consider first the $g=3$ plots in Fig. 7(a). The succession of the peaks of $\eta_m^{3,3}(r)$ follows clearly the ascending order of $m$. For the $g=4$ LDPs a similar, essentially ascending, succession is found for the peaks of $\eta_m^{4,4}(r)$, shown in Fig. 7(b), except that the distributions for the first and second neighbours peak at nearly the same distance. The same succession is exhibited by the short distance behaviour of these functions (inset in Fig. 7(b)) where in fact the $m=2$ curve appears to slightly exceed the $m=1$ and both to be well above the $m=3$ curve which, in turn, is well above the $m=4$ and $m=5$ curves. Thus the proximity of the outer shell sites follows the neighbour index succession $m=1\approx 2,3,4,5$. The plots for $g=5$, shown in Fig. 7(c), exhibit a quite different succession of proximity: the peak of the second neighbour sites ($m=2$) is at larger distance than that of the fourth neighbour sites ($m=4$), with the distribution of the latter peaking at nearly the same distance as the first neigbours. In fact, the average proximity of a site to its neighbours, from first to fifth, follows the succession $m=1,4,3,2,5$. This succession holds also for the short distance behaviour of $\eta_m^{5,5}(r)$ depicted in the inset of Fig. 7(c).

These results are in accord with the existence of a twisted disposition of the dendra along the backbone, with the twist angle step being roughly $120°$ for the $g=4$ LDP and $90°$ for the $g=5$. In particular, they support a picture whereby the dendra grow from sites located at regular distances along the backbone but expand along different directions in order to use all the available free space around the backbone. For symmetry reasons, it is expected that these directions are not random but follow some specific repetition scheme according to which, for the $g=5$ LDP, every fourth neighbor dendron is brought in register with the initial one, i.e. for every four dendrons we have the formation of a complete pitch of a helix. It is obvious that in this case the torsion angle formed between first neighbor dendra should be around one fourth of a full rotation. For the $g=4$ supermolecule, the same phenomenon is observed, but in this case it is the third neighbor dendron that is brought in register with the starting one, thus forming a helix of torsion-angle step of around one third of a full rotation.

## IV. ELONGATION, STIFFNESS AND FORM-FACTOR RESULTS

The torsion angle distributions, the backbone curling angle distributions and the topologically specific correlation functions described in the previous section reveal the formation of persistent helical structures on going from $g=4$ to $5$. Here we examine how this transition is reflected on other structural properties of the LDPs.

We consider first the equilibrium probability distribution function of the end-to-end distance $R_{1,N}$ of the backbone. For a LDP of generation $g$ this function is denoted by $P(R_{1,N};g)$. The results of our calculation for these probability distributions are plotted in Fig. 8. It is clear from these plots that while the low generation distributions are broad and exhibit several peaks, the distributions for g=4 and 5 consist a single, relatively narrow peak. Moreover, as shown in Fig. 8, these high generation distributions can be fairly accurately approximated by single Gaussian curves, in contrast to the $g=0$ to 3 for which a Gaussian fit is inadequate qualitatively.

We have used the distributions $P(R_{1,N};g)$ of Fig. 8 for the evaluation the average end-to-end length of the backbone, $R_{e-e}\equiv\langle R_{1,N}\rangle$. We have also evaluated the second moments $\left(\langle R_{1,N}^2\rangle-R_{e-e}^2\right)$ of these distributions, which provide a measure of the effective compliance of the backbone to extensions of its length. In particular, within the single Gaussian approximation of the $P(R_{1,N};g)$ distribution, the LDP behaves as an ideal spring when a pair of sufficiently small forces (directed in opposite directions along the end-to-end vector) is applied at its two ends. In that case, the respective spring constant $k_{el}$ is inversely proportional to the second moment of the end-to-end distribution,[23] namely $k_{el}\sim 1/\left(\langle R_{1,N}^2\rangle-R_{e-e}^2\right)$, while the natural length of the spring corresponds to the first moment of this distribution, namely $R_{e-e}$. Fig. 9 shows the results of our calculation for $R_{e-e}$ and $\left(\langle R_{1,N}^2\rangle-R_{e-e}^2\right)^{-1}$ as a function of generation $g$. Notably the inverse second moment plot (Fig. 9(b)) exhibits an increase by two order of magnitude from $g=3$ to $g=5$, while the respective variation of $R_{e-e}$ plot (Fig. 9(a)) is marginal. Thus the effective stiffness of the LDP undergoes a dramatic increase above $g=3$.

The results exhibit some qualitative deviations from the results obtained in Ref. 4, where the average end-to end length was found to increase with $g$ up to $g=4$ and then to decrease for $g=5$ while the decrease of the second moments with $g$ was found to be much milder (a factor of about 1.5 was found there on going form $g=0$ to $g=5$, compared to the factor of 70 appearing in Fig. 9(b)). These deviations stem from the different treatment of the backbone bond-stretching in the two calculations: while here we have allowed, through the $u^{(BL)}(l_{ss'})$ terms of the potential of Eq. (1), for free bond stretching within the range $(0.67 \pm 0.09)D$, the bonds in Ref. 4 were taken to have fixed lengths, thus producing a backbone with much higher intrinsic stiffness and therefore limited margins of further stiffening with increasing dendron generation. In contrast, by starting out with an intrinsically more stretchable backbone, as done in the present work, the effect of packing-induced stiffening for large $g$ can be magnified and clearly separated from the intrinsic stiffness of the backbone. On the other hand, this comparison reveals that the calculation of the generation dependence of the leading moments of the probability distribution function $P(R_{1,N}; g)$ is sensitive to the modeling of the bond length dependent terms $u^{(BL)}(l_{ss'})$ of the LDP potential energy of Eq. (1).

Finally, we consider the mean correlation function defined as

$$\eta(r) \equiv \frac{2}{N_{site}(N_{site}+1)} \left\langle \sum_{\{s,s'\}} \delta(r - r_{s,s'}) \right\rangle. \quad (7)$$

The summation in Eq. (7) runs over all the possible pairs of molecular sites and $N_{site}$ denotes the total number of sites. Thus $\eta(r)$ gives the probability of finding any pair of sites, irrespectively of their specific connectivity within the LDP, at separation $r$. Accordingly, the calculation of $\eta(r)$ provides a direct way to determine the density-density autocorrelation function and the form factor[23,24] of a single LDP.

We have evaluated in our simulations the correlation functions $\eta(r)$ for the LDPs of generation $g=0$ to 5. The average mass distributions described by $\eta(r)$ for generations $g=4$ and 5 are plotted in Fig. 10. There are four peaks in these plots that can be identified easily. For $g=4$, the first small peak at $0.78D$ corresponds to the distance between two neighboring sites on the backbone. The second peak, at 1.10D corresponds to the distance between bonded united atoms of the dendra. The third peak is at 2.05D, which is the characteristic distance between second-neighbor sites of the dendra. The region in which the highest and broadest peak is observed corresponds to the range of distances between any two non-bonded pairs of united atoms of the macromolecule. A similar assignment holds for the peaks in the $g=5$ plot, where the peaks associated with non-bonded pairs are now somewhat broader and the distribution is shifted to longer distances, as expected for a more extended structure. This sort of behaviour has been observed in small-angle X-ray and neutron scattering experiments.[25] It is apparent, however, from the plots of Fig. 10, that $\eta(r)$ is not directly responsive to the formation of helical structures, in contrast to the topologically specific correlation functions introduced in Sec. III, whose qualitative trends change as the helical structures set in.

## V. CONCLUSIONS AND DISCUSSION

Our molecular simulations of LDPs, based on the Metropolis Monte Carlo method and using excluded volume interactions among united atom molecular segments, show clear signs of the onset of helical structures at dendritic generation $g=4$ and of their dramatic enhancement at $g=5$.

The helical structures are manifested by a twisting of the relative disposition of successive dendra along the backbone accompanied by a slight curling of the backbone. The helical pitch, which is common for the twisting and the curling, depends on the dendritic generation. In particular, the pitch is found to be three and four repeat-units long for the $g=4$ and the $g=5$ LDPs respectively. This corresponds to respective torsion-angle increments of 120° and 90° on moving along the backbone from one dendron grafting site to the next. The appearance of helical structures is also accompanied by an abrupt stiffening of the LDP with respect to elongations.

The helical structures exhibit defects, consisting of domains with twisting in the reverse sense to the dominant twisting. For $g=4$, such defects can appear and disappear, propagate along the backbone, merge to form larger domains and even lead to complete reversal of the dominant twisting sense. For $g=5$, however, the appearance of defects depends on the starting conditions and, if present, these defects remain fixed over very long simulation runs.

The appearance of macro-chiral structures in these chemically achiral LDPs is attributed to the strong packing constraints imposed on the dendritic mass as the generation increases beyond $g=3$, while keeping the grafting intervals along the backbone constant. Such packing-induced chirality is macroscopically symmetric with respect to the reversal of the helical handedness i.e. each state of the LDP has the same free energy with the state (enatiomer) obtained from it by changing the signs of all the twisting and curling dihedral angles. This symmetry-imposed bistability is of course lost if some intrinsically chiral groups are introduced into the chemical structure of the LDPs. In such case, the free energy of one of the two enatiomeric states will have lower energy and will thus be favoured thermodynamically over the other. If, moreover, the intrinsic chirality can be externally controlled (e.g. photo-induced) the LDP can in principle exhibit tunable chirality. However, our present calculations do not allow any realistic estimates of the chirality switching speed in such systems.

The helical structures identified in the present study are of short pitch (3-4 repeat units). It should be stressed, however, that the model LDPs we have simulated are relatively short (~ 20 repeat units) and therefore our results do not exclude the possibility of the appearance of a second type of helical structures, generated by longer pitch twisting of the dendra and/or curling of the backbone, that would be superposed to the short pitch structures.

Finally, it is worth noting that while the results described in this study are based on hard-body repulsive interactions, the qualitative physical picture persists for soft repulsive models of the potential. This has been confirmed by repeating a selected set of simulations with the hard body interactions replaced by soft ones, specifically, soft sphere potentials and repulsive potentials varying according to an inverse power law of the intersegmental distance. The quantitative results obtained in this way are sensitive to the parameterization of the soft potential and obviously temperature dependent. In particular, the persistent defects obtained for certain starting configurations of the g=5 LDPs in the hard body model are invariably found to unlock on raising the temperature in the soft repulsion model. Nevertheless, the appearance of the helical structures above a certain generation and below a certain temperature appears to be a common feature of these simulations that stands beyond the details of the parameterization used for the repulsive potential.

## ACKNOWLEDGMENTS

This work was supported by the Greek Secretariat for Research and Technology (GGET) by a grant (PENED 01ED529).

**FIGURE CAPTIONS**

**FIG. 1**. The coarse grained modeling of LDPs employed in the simulations, shown here for a first generation model LDP. dendronized polymer. Only the phenyls are considered as united atoms. Monomer of generation g=1 is shown here.

**FIG. 2**. Backbone unit vectors $\mathbf{e}_i$ and dendron grafting unit vectors $\mathbf{u}_i$ used for the definition of the torsion angles $\phi_i$, the backbone bending angles $\theta_i$ and curling angles $\psi_i$. The spheres describe the phenyls of the backbone and the cones describe the grafted dendra.

**FIG. 3**. Calculated probability distributions $p^{(t)}(\phi;g)$ of the torsion angle $\varphi$, defined in Eq. (3) for LDPs of generations g=0 to 5. Due to the physical coincidence of the $\phi = \pm 180°$ angles, the values $p^{(t)}(180°, g)$ and $p^{(t)}(-180°, g)$ are identical.

**FIG. 4**. Snapshots illustrating equilibrium structures of the 20-monomer LDP with g = 0 to g = 5 dendra attached. Colour code: backbone sites are marked with green colour; quartets of successive dendrons are coloured with blue, red, yellow and magenta.

**FIG. 5.** (a) Calculated backbone bending-angle distributions $p^{(b)}(\cos\theta, g)$ and (b) calculated averages of the generation dependence of the backbone bending-angle averages $<\cos\theta>$.

**FIG. 6.** Calculated mean probability distribution $p^{(c)}(\psi;g)$ of the backbone-curling dihedral angle $\psi$ for the LDPs of generation (a) g=2 and (b) g=5.

**FIG. 7**. Plots of the calculated topologically specific radial correlation functions $\eta_m^{g_t, g_t'}(r)$, defined in Eq. (6), for pairs of sites belonging to the outer topological shells ($g_t = g_t' = g$) of dendra grafted at the m=1,2,3,4 and 5 neighbour positions for LDPs of generations g=3, g=4 and g=5. The short distance behaviour of the correlation function is magnified in the insets of g=4 and g=5.

**FIG. 8**. Calculated probability distribution function $P(R_{1,N};g)$ of the end-to-end distance $R_{1,N}$ of the backbone for LDPs of generations g=0 to 5. The solid curves represent Gaussian best fits to the calculated distributions. The Gaussian best fit parameters, in the notation $G(x) = \dfrac{e^{-(x-x_0)^2/2\sigma^2}}{\sigma\sqrt{2\pi}}$, are indicated.

**FIG. 9**. (a) Generation dependence of the average end-to-end length of the LDPs, scaled by the maximum achievable length of the backbone ($R^{\max}$). (b) Generation dependence of the inverse second moments $\left(\langle R_{1,N}^2 \rangle - R_{e-e}^2\right)^{-1}$ of the $P(R_{1,N};g)$ distribution, scaled by the inverse second moment of the $g=0$ LDP.

**FIG. 10**. Plots of the calculated average number density around a site, as described by the radial correlation function $\eta(r)$ defined in Eq. (7), for the LDPs of generations $g=4$ and 5. The short distance behaviour of $\eta(r)$ is magnified in the inset.

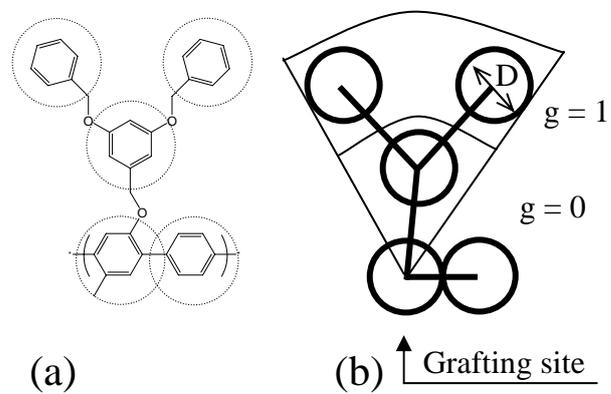

D.K. Christopoulos et al., **FIGURE 1**

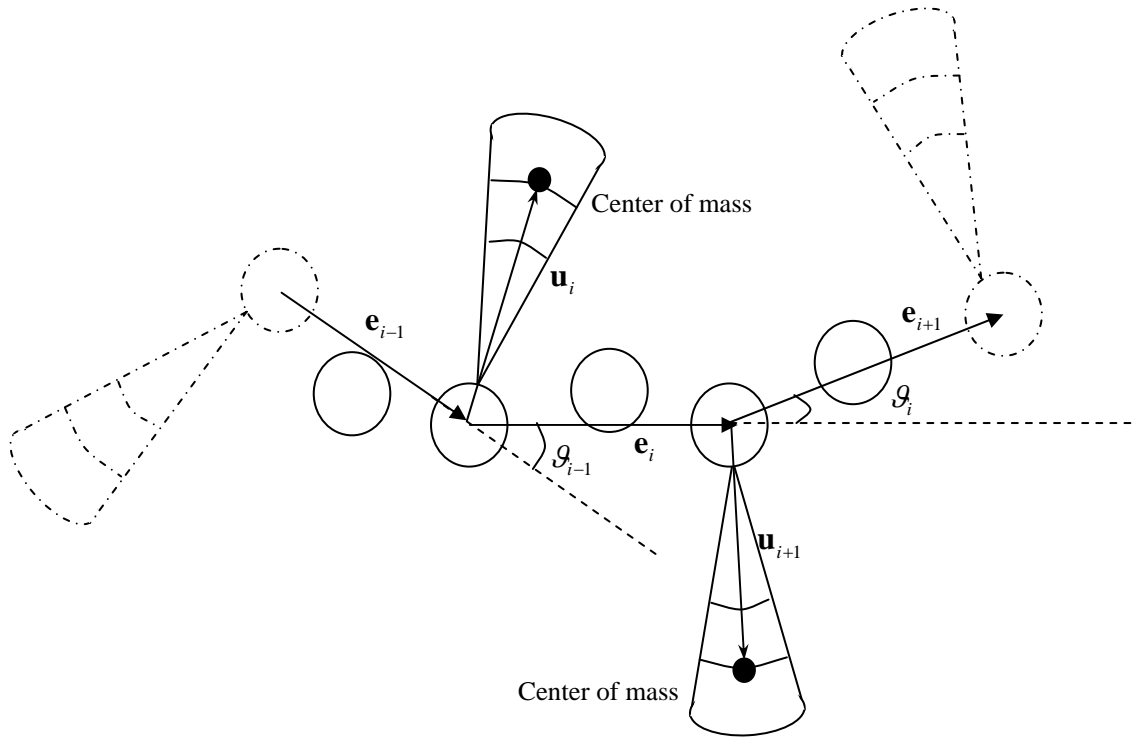

D.K. Christopoulos et al., **FIGURE 2**

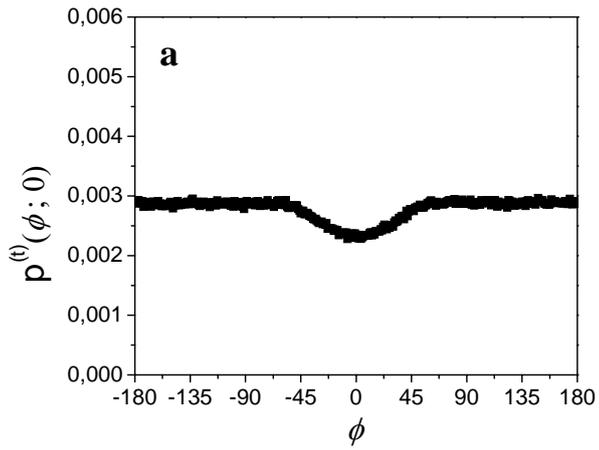
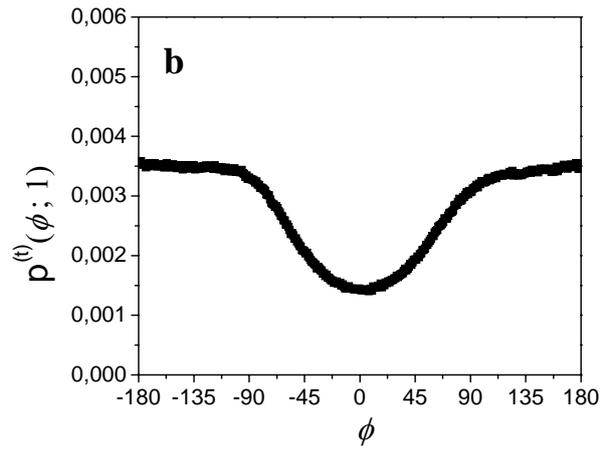
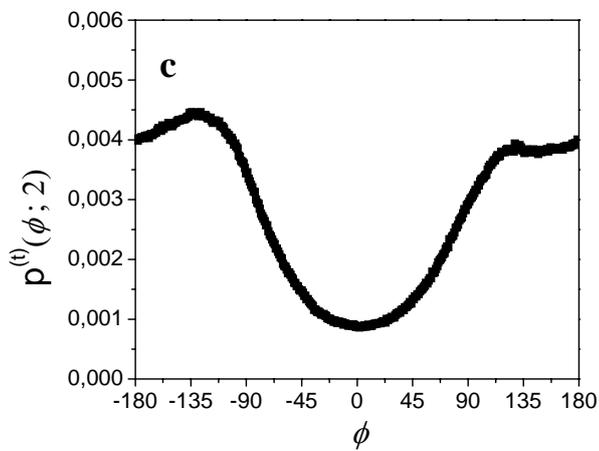
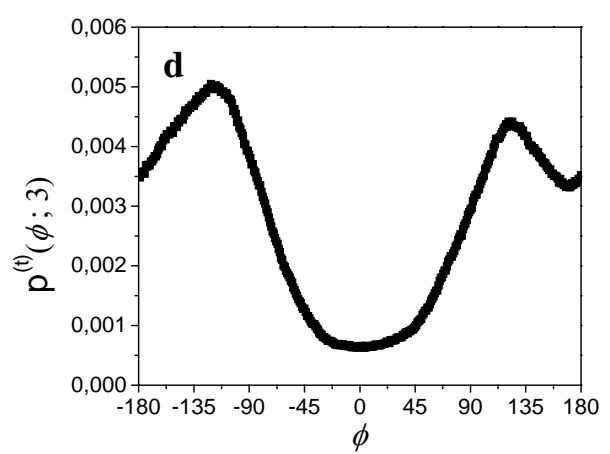
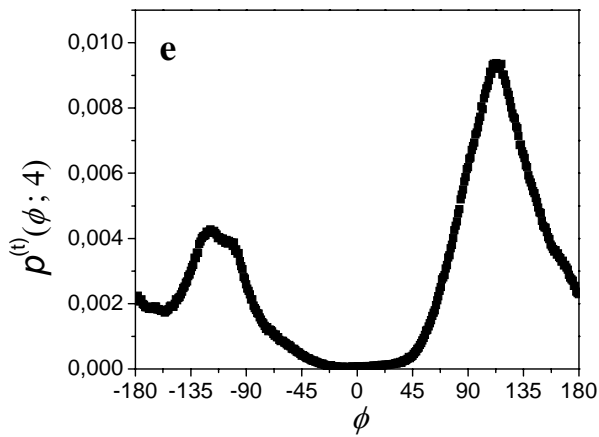
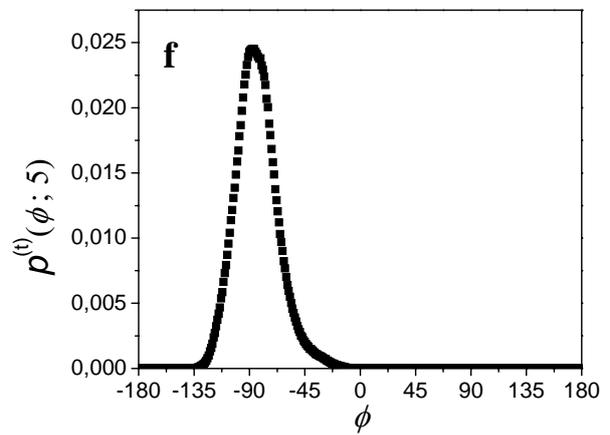

D.K. Christopoulos et al., **FIGURE 3**

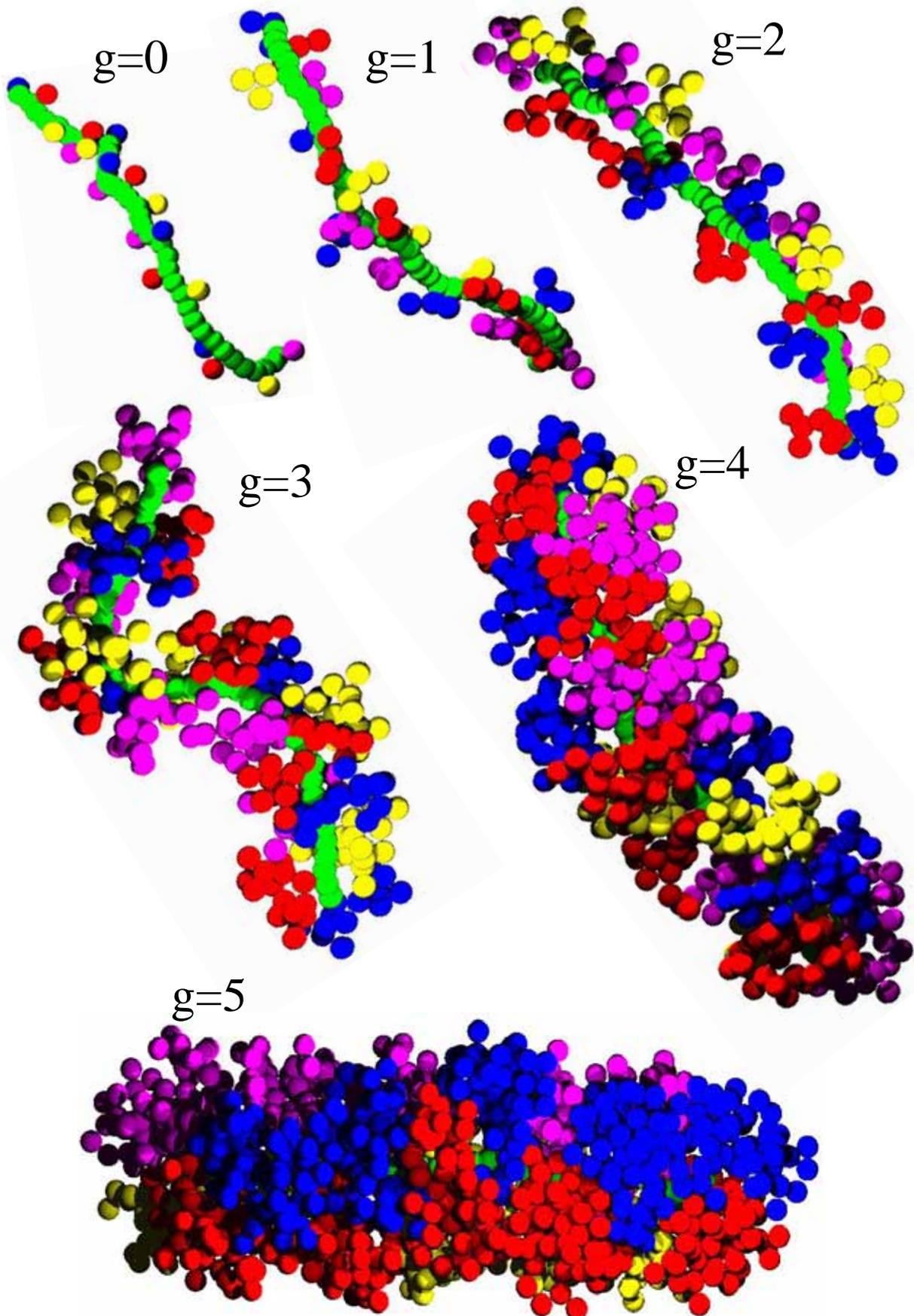

D.K. Christopoulos et al., **FIGURE 4**

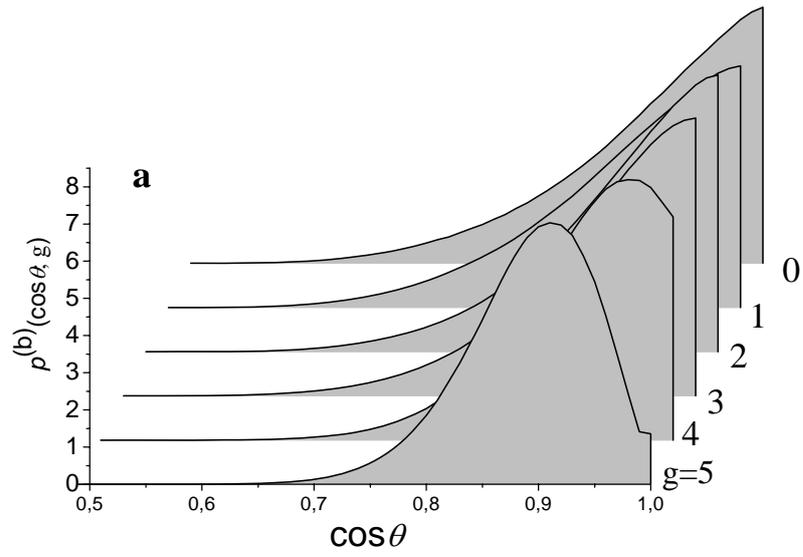

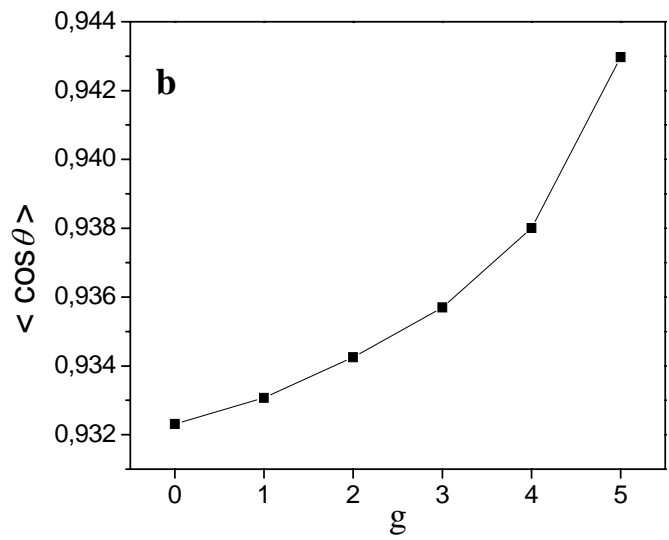

D.K. Christopoulos et al., **FIGURE 5**

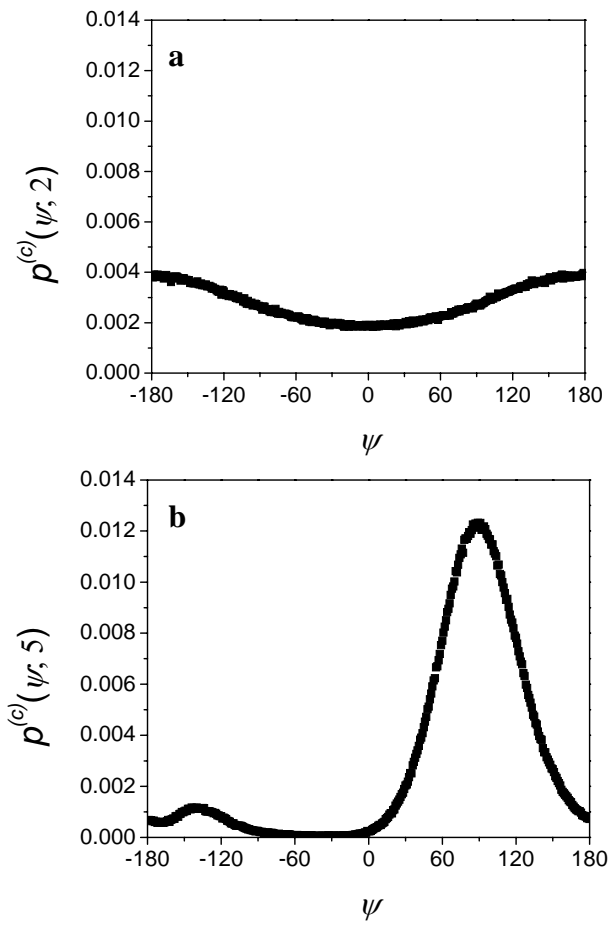

D.K. Christopoulos et al., **FIGURE 6**

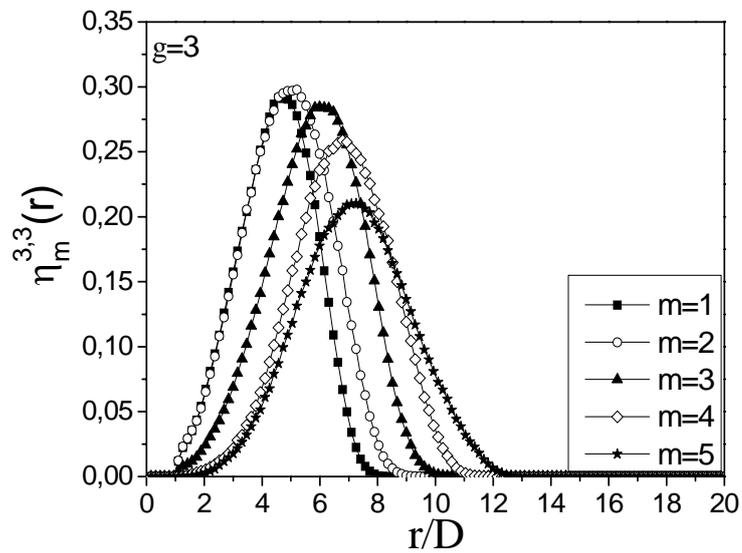

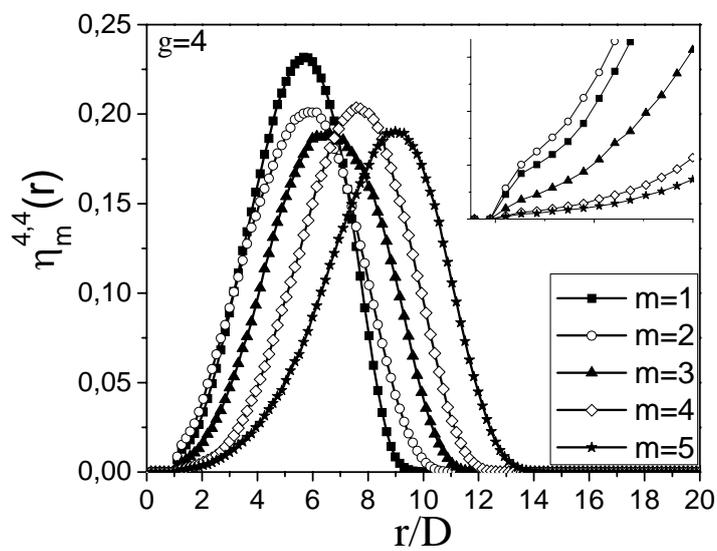

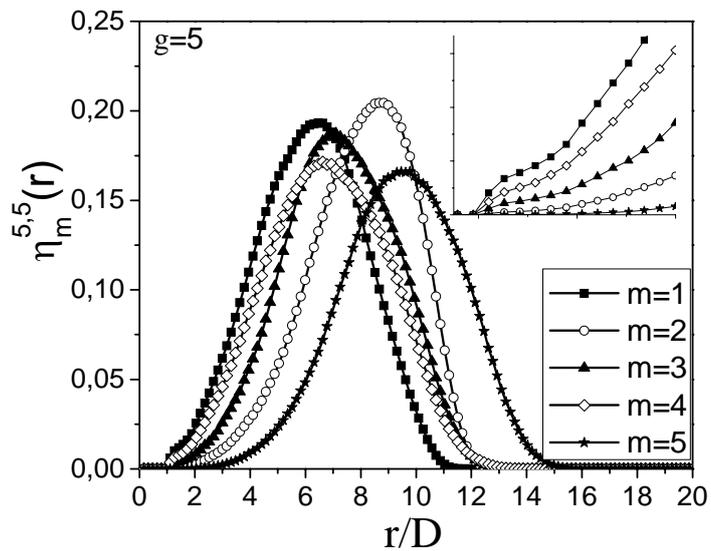

D.K. Christopoulos et al., **FIGURE 7**

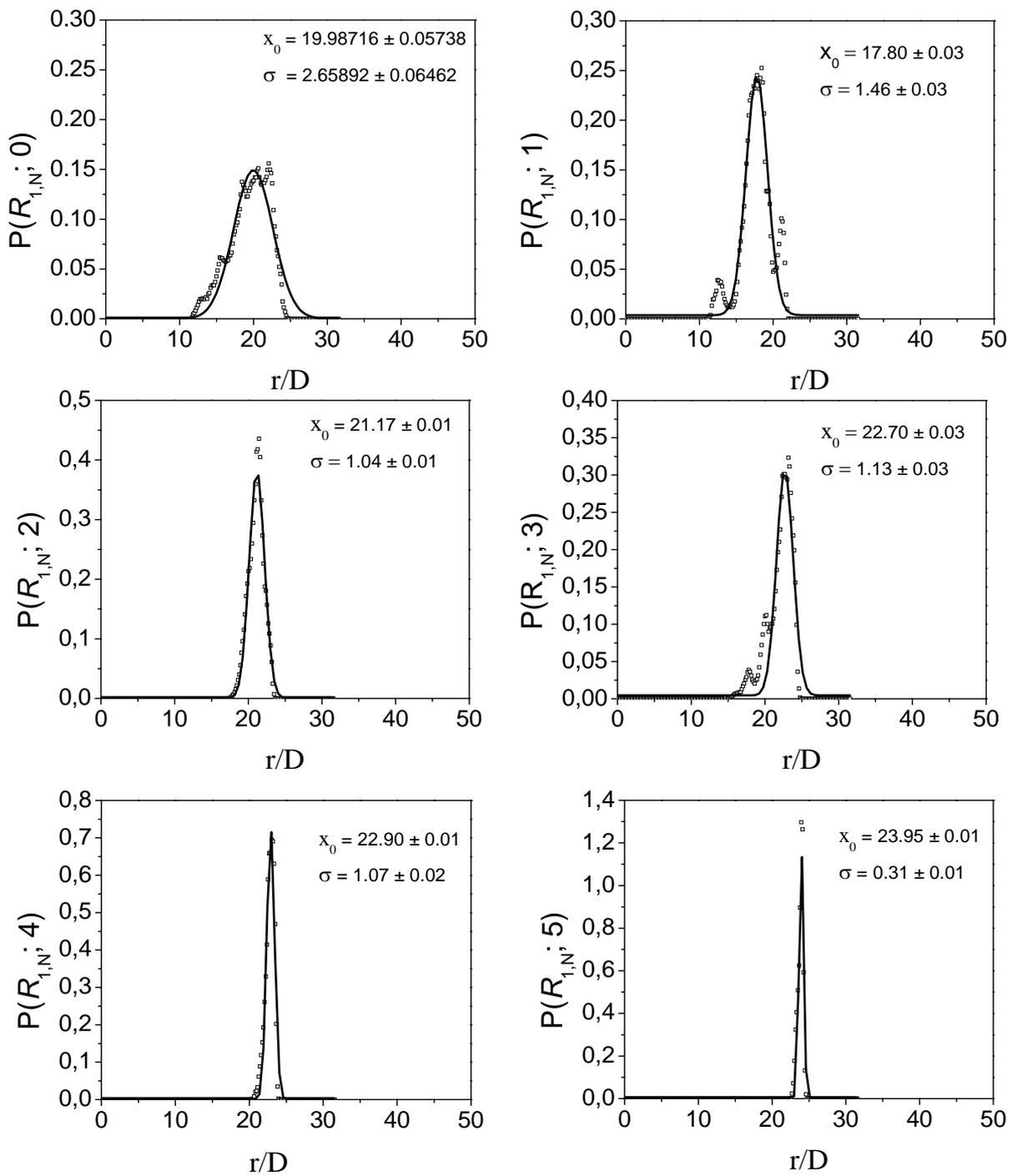

D.K. Christopoulos et al., **FIGURE 8**

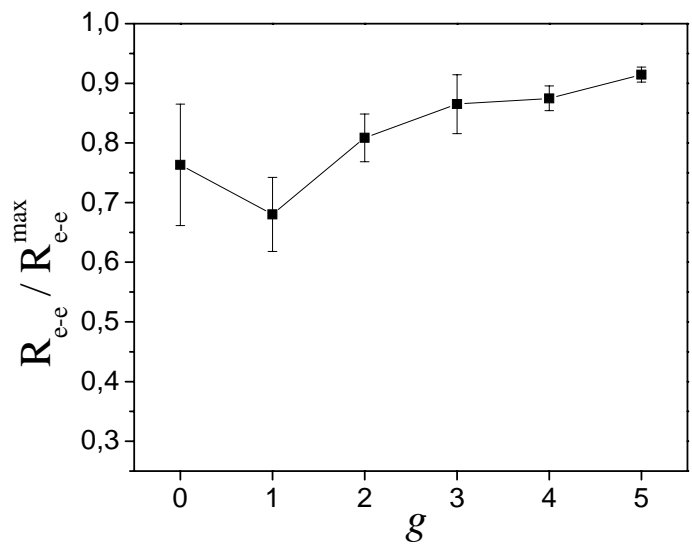

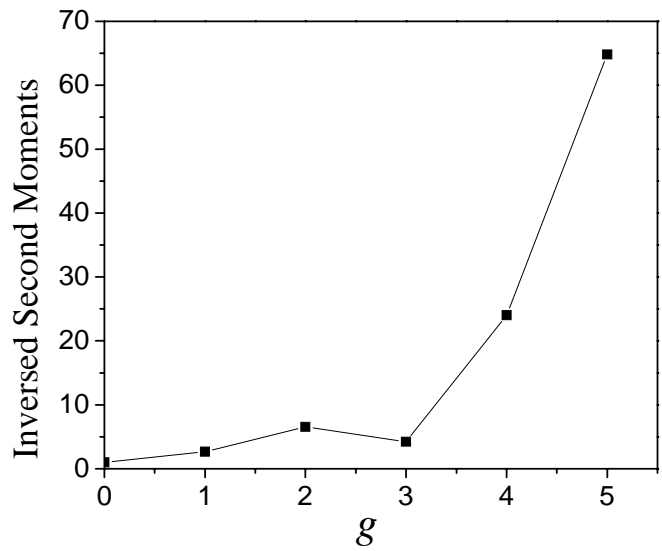

D.K. Christopoulos et al., **FIGURE 9**

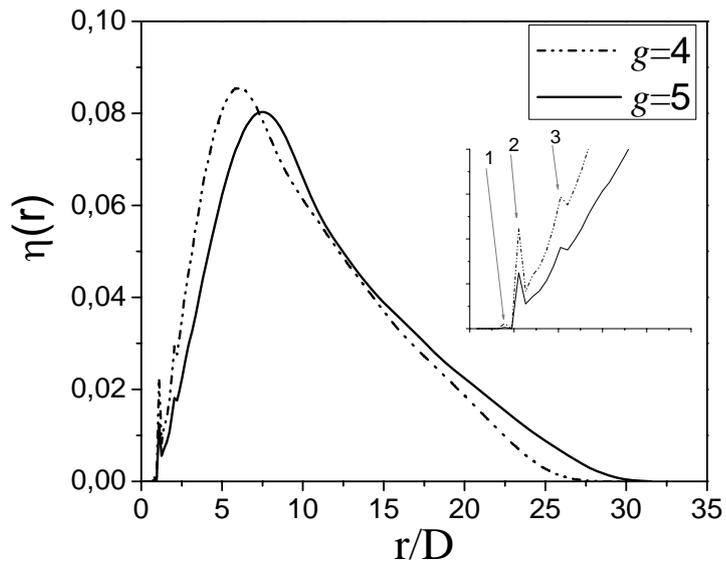

D.K. Christopoulos et al., **FIGURE 10**